\documentclass[aps,prb,twocolumn,showpacs,preprintnumbers,amsmath,amssymb,superscriptaddress]{revtex4}%

\usepackage{graphicx}%
\usepackage{dcolumn}
\usepackage{amsmath}
\usepackage{color}
\usepackage{bm}

\begin{document}


\title{Correlation between the transition temperature and the superfluid density in BCS
superconductor NbB$_{2+x}$}

\author{R.~Khasanov}
 \affiliation{Physik-Institut der Universit\"{a}t Z\"{u}rich,
Winterthurerstrasse 190, CH-8057 Z\"urich, Switzerland}
\author{A.~Shengelaya}
 \affiliation{Physics Institute of Tbilisi State University,
Chavchavadze 3, GE-0128 Tbilisi, Georgia } 
\author{A.~Maisuradze}
 \affiliation{Physik-Institut der Universit\"{a}t Z\"{u}rich,
Winterthurerstrasse 190, CH-8057 Z\"urich, Switzerland} 
\author{D. Di~Castro}
\affiliation{Physik-Institut der Universit\"{a}t Z\"{u}rich,
Winterthurerstrasse 190, CH-8057 Z\"urich, Switzerland}
\affiliation{INFM-Coherentia and Dipartimento di Fisica, Universita'
di Roma "La Sapienza", P.le A. Moro 2, I-00185 Roma, Italy}
\author{R.~Escamilla}
\affiliation{Instituto de Investigaciones en Materiales, Universidad
Nacional Aut\'{o}noma de M\'{e}xico, 04510 M\'{e}xico D.F.,
M\'{e}xico}
\author{H.~Keller}
 \affiliation{Physik-Institut der Universit\"{a}t Z\"{u}rich,
Winterthurerstrasse 190, CH-8057 Z\"urich, Switzerland}

\begin{abstract}
The results of the muon-spin rotation experiments on BCS
superconductors NbB$_{2+x}$ ($x = 0.2$, 0.34) are reported. Both
samples, studied in the present work, exhibit rather broad
transitions to the superconducting state, suggesting a distribution
of the volume fractions with different transition temperatures
($T_c$)'s. By taking these distributions into account, the
dependence of the inverse squared zero-temperature magnetic
penetration depth ($\lambda_0^{-2}$) on $T_c$ was reconstructed for
temperatures in the range $1.5$~K$\lesssim T_c \lesssim 8.0$~K.
$\lambda_0^{-2}$ was found to obey the power law dependence
$\lambda_0^{-2}\propto T_c^{3.1(1)}$ which appears to be common for
some families of BCS superconductors as, {\it e.g.}, Al doped
MgB$_2$ and high-temperature cuprate superconductors as underdoped
YBa$_2$Cu$_3$O$_{7-\delta}$.

\end{abstract}


\maketitle

\section{Introduction}

The search for universal correlations between physical variables,
such as transition temperature, magnetic field penetration depth,
electrical conductivity, energy gap, Fermi energy {\it etc.} may
provide hints towards a unique classification of different
superconductors. In particular, establishing of such correlations
may help to understand the phenomenon of superconductivity, that is
observed now in quite different systems, such as simple metals and
alloys, fullerenes, molecular metals, cuprates, cobaltites, borides
{\it etc}. Among others, there is a correlation between the
transition temperature ($T_c$) and the inverse squared
zero-temperature magnetic field penetration depth
($\lambda_0^{-2}$), that generally relates to the zero-temperature
superfluid density ($\rho_s$)  in terms of
$\rho_s\propto\lambda_0^{-2}$. In various families of underdoped
high-temperature cuprate superconductors (HTS)'s there is the
empirical relation $T_c\propto\rho_s\propto\lambda_0^{-2}$, first
identified by Uemura {\it et al.}\cite{Uemura89,Uemura91} It was
recently shown, however, that for HTS's with highly reduced $T_c$'s
the direct proportionality between $T_c$ and $\lambda_0^{-2}$ is
changed to a power law kind of dependence with
$(T_c)^n\propto\lambda_0^{-2}$ ($n$ is the power law exponent). In
experiments on highly underdoped YBa$_2$Cu$_3$O$_{7-\delta}$ Zuev
{\it et al.}\cite{Zuev05} obtained $n = 2.3$, Liang and coworkers
reported $n = 1.6$,\cite{Liang05} while Sonier {\it et
al.}\cite{Sonier07} found $n=2.6-3.1$. In molecular superconductors
Pratt and Blundell obtained $n = 2/3$.\cite{Pratt05} From the
theoretical point of view, it was shown that in systems obeying 2D
or 3D quantum superconductor to insulator transition, $n\equiv 1$ or
$n\equiv 2$,
respectively.\cite{Kim91,Schneider00,Schneider04,Schneider07}

It should be emphasized, however, that the relation between $T_c$
and $\lambda_0^{-2}$ is not yet established for BCS superconductors
and still awaits to be explored.
A good candidate to search for such a relation could be NbB$_{2+x}$.
The superconductivity in NbB$_{2+x}$, similar to MgB$_2$, is most
likely mediated by phonons. It is confirmed in nuclear magnetic
resonance (NMR)\cite{Kotegawa02} and tunnelling
experiments,\cite{Takasaki04,Ekino04} as well as by recent
calculations of the elastic properties.\cite{Regalado07} Moreover,
muon-spin rotation ($\mu$SR) experiments suggest that the
superconducting gap is isotropic.\cite{Takagiwa04} As is shown in
Refs.~\onlinecite{Escamilla04} and \onlinecite{Yamamoto01} the
superconductivity in NbB$_2$ can be induced by either increasing
boron or decreasing niobium content, while the parent NbB$_2$
compound is not superconducting at least down to 2~K. The transition
temperature was found to reach the maximum value of $9.2$~K and
$9.8$~K for Nb$_{0.9}$B$_2$ and NbB$_{2.34}$,
respectively.\cite{Escamilla04,Yamamoto01} This offers a possibility
to study the relation between $T_c$  and $\lambda_0^{-2}$ as a
function of boron and/or niobium content. In the present study the
temperature dependence of the magnetic field penetration depth was
measured for two NbB$_{2+x}$ samples with $x = 0,2$ and 0.34 by
means of transverse-field muon-spin rotation technique. It was found
that in both samples the distribution of the superconducting volume
fractions with different $T_c$'s can be well approximated by a
Gaussian distribution. The mean values of the superconducting
transition temperature ($T^m_c$) and the width of the distribution
($\Delta T_c$) were found to be $T^m_c= 6.02(3)$~K, $\Delta
T_c=0.96(2)$~K for NbB$_{2.34}$ at $\mu_0H = 0.1$~T, and
$T^m_c=3.40(4)$~K, $\Delta T_c = 1.06(2)$~K for NbB$_{2.2}$ at
$\mu_0H = 0.05$~T. Within the model, developed for a granular
superconductor of moderate quality, we reconstruct the dependence of
the zero-temperature superfluid density
$\rho_s\propto\lambda_0^{-2}$ on the transition temperature $T_c$.
It was found that in the range of $1.5$~K$\lesssim T_c \lesssim
8.0$~K $\lambda_0^{-2}$ follows a power law dependence, rather than
a linear dependence reported by Takagiwa {\it et
al.},\cite{Takagiwa04} with $\lambda_0^{-2}\propto T_c^{3.1(1)}$.
The value of the power law exponent 3.1(1) agrees rather well with
$n=2.6-3.1$ reported by Sonier {\it et al.}\cite{Sonier07} for
underdoped HTS's YBa$_2$Cu$_3$O$_{7-\delta}$.

The paper is organized as follows. In Sec.~\ref{subsec:Theoretical
background-calculations} we describe the model used to obtain the
temperature dependence of the magnetic field penetration depth for a
granular superconductor having a certain distribution of the
superconducting volume fractions. The distributions of the local
magnetic fields, calculated within the framework of this model, are
presented in Sec.~\ref{subsec:Theoretical background-simulations}.
In Sec.~\ref{sec:experimental} we describe the sample preparation
procedure and details of the muon-spin rotation and magnetization
experiments. Sec.~\ref{sec:results_and_discussions} comprises
studies of the magnetic penetration depth in NbB$_{2.2}$ and
NbB$_{2.34}$ superconductors. The conclusions follow in
Section~\ref{sec:conclusions}.

\section{Theoretical background}\label{sec:theoretical
background}

\subsection{Magnetic penetration depth in a granular superconductor of
moderate quality} \label{subsec:Theoretical
background-calculations}

In this section we describe the model applied to calculate
temperature dependence of the magnetic field penetration depth
$\lambda$ in a granular superconductor of moderate quality by using
the $\mu$SR data. We based on a general assumption that $\lambda$
can be obtained from the second moment of the local magnetic field
distribution [$P(B)$] inside the superconducting sample in the mixed
state measured directly in $\mu$SR experiments .
The model uses following assumptions: (i) The superconducting grains
are decoupled from each other. (ii) Each $i-$th grain is a
superconductor with a certain value of the transition temperature
($T^i_c$) and the zero-temperature magnetic penetration depth
($\lambda^i_0$). (iii) The zero-temperature superconducting gap
($\Delta_0$) scales with $T_c$ in agreement with the well-known
relation $\Delta_0/k_BT_c = const$,\cite{Tinkham75} implying that
the ratio $R = \Delta_0^i/k_BT_c^i$ is the same for all the grains.
Note that a linear decrease of both superconducting energy gaps
($\Delta^\sigma$ and $\Delta^\pi$) with decreasing $T_c$ was
observed recently by Gonelli {\it et al.}\cite{Gonnelli06} in Mn
doped MgB$_2$. The similar linear $\Delta_0$ {\it vs.} $T_c$ scaling
was reported by Khasanov {\it et al.}\cite{Khasanov04} for
RbOs$_2$O$_6$ BCS superconductor.

Let us first define variables and functions used within the model.
Function $\omega(t)$ describes the distribution of the
superconducting volume fractions with different transition
temperatures $T_c$'s. It is defined so that the volume fraction of
the sample, having transition temperatures in the range between
$T_c^i$ and $T_c^j$, is obtained as $\int_{T_c^i}^{T_c^j}
\omega(t)dt$. %
The distribution of the local magnetic fields in the $i-$th grain is
described by the function $P^i(B)$, which for ideal superconductor
has rather asymmetric shape [see Fig.~\ref{fig:simulations}~(a)].
Function $f(t)$ describes dependence of the inverse squared magnetic
penetration depth $\lambda^{-2}$ at $T = 0$ on $T_c$
[$(\lambda_0^i)^{-2} = f(T_c^i)$]. The dependence of $\lambda^i$ on
temperature was assumed to follow the standard equation for weak
coupled BCS superconductor [see, {\it e.g.}, Eq.~(2-111) in
Ref.~\onlinecite{Tinkham75}]:
\begin{equation}
[\lambda^i(T)]^{-2} = (\lambda^i_0)^{-2}s(T,\Delta_0^i)=
f(T_c^i)s(T,k_BR\cdot T_c^i),
 \label{eq:BCS-weak-coupled}
\end{equation}
with the temperature dependent part
\begin{equation}
s(T,\Delta_0^i)= 1+ 2\int_{\Delta^i(T)}^{\infty}\left(\frac{\partial
F}{\partial E}\right)\frac{E}{\sqrt{E^2-\Delta^i(T)^2}}\  dE.
\nonumber
\end{equation}
Here, $F=[1+\exp(E/k_BT)]^{-1}$ is  the Fermi function,
$\Delta^i(T)=\Delta_0^i \tilde{\Delta}(T/T_c^i)$ represents the
temperature dependence of the energy gap,  and
$\Delta^i_0=2Rk_BT_c^i$ denotes the zero temperature value of the
superconducting gap. For the normalized gap
$\tilde{\Delta}(T/T_c)$ the values tabulated in
Ref.~\onlinecite{Muhlschlegel59} were used.
Finally, the temperature dependence of the total second moment of
the local magnetic field distribution in the superconducting sample
can obtained as:
\begin{equation}
\langle \Delta B^{2}\rangle^{tot} =\frac{\sigma^2}{\gamma_\mu^2}
=\int_T^\infty \int_0^\infty \omega(t) P(B',t)(B'-B^{m})^2dtdB'.
 \label{eq:B_tot-full}
\end{equation}
Here, $\gamma_\mu = 2\pi\times135.5342$~MHz/T is the muon
gyromagnetic ratio, $\sigma^{2}$ is the second moment of the
$\mu$SR line and $B^m$ is the mean internal field inside the
superconducting sample.

As will be shown later, in transverse-field $\mu$SR experiments
one obtains directly: (i) the distribution of the superconducting
volume fractions with different $T_c$'s [$\omega(T_c)$], (ii) the
temperature dependence of the mean internal field $B^m(T)$, and
(iii) the temperature dependence of the second moment of the
$\mu$SR line $\sigma^2(T)$. By substituting $\omega(T_c)$ and
$B^m(T)$ to the Eq.~(\ref{eq:B_tot-full}) and then fitting it to
the experimental $\sigma(T)$ data, one should be able to obtain
the distribution of the zero-temperature superfluid density
$\rho_s\propto\lambda_0^{-2}$ as a function of the transition
temperature $T_c$ and the ratio $R=\Delta_0/k_BT_c$. In order to
do that one needs, however, to calculate $P^i(B)$ distributions
which depend on the applied field ($B_{ex}$), $\lambda^i$, and the
coherence length ($\xi^i$) in a nontrivial way. This makes fit of
Eq.~(\ref{eq:B_tot-full}) to the experimental data extremely
difficult.
The analysis can be simplified by assuming that each $P^i(B)$
follows the Gaussian distribution and is determined, therefore, by
the only two parameters: the second moment of the Gaussian line
\begin{equation}
\langle \Delta B^{2}\rangle =
(\sigma/\gamma_\mu)^2=G^2(b)\cdot(\lambda)^{-4},
 \nonumber
\end{equation}
and the internal field of the $i-$th grain ($B^i$). Here, $b=
B/B_{c2}$ is the reduced magnetic field ($B_{c2}$ is the second
critical field) and the $G(b)$ is the proportionality coefficient
between $\sigma/\gamma_\mu=\sqrt{\langle \Delta B^{2}\rangle}$ and
$\lambda^{-2}$ which can be obtained by means of Eq.~(13) from
Ref.~\onlinecite{Brandt03} as:
\begin{equation}
G(b)=0.172\frac{\Phi_0}{2\pi}(1-b)[1+1.21(1-\sqrt{b})^3],
  \label{eq:G(b)}
\end{equation}
($\Phi_0$ is the magnetic flux quantum). Here we also take into
account that within Ginzburg-Landau theory $\xi=\sqrt{\Phi_0/2\pi
B_{c2}}$. The internal field of the $i-$th grain was calculated
within the London approximation modified by Brandt:\cite{Brandt03}
\begin{eqnarray}
B^i&=&B_{ex}-(1-D^i)\frac{\Phi_0}{8\pi\cdot
(\lambda^i)^2}\ln[g(b^i)]
\nonumber \\
    &\simeq&B_{ex}-(1-D^m)\frac{\Phi_0}{8\pi}f(T_c^i)s(T,k_BRT_c^i)\ln[g(b^m)]
 \label{eq:Bi}
\end{eqnarray}
where
\begin{equation}
 g(b) = 1 + \frac{1-b}{b}(0.357+2.890b-1.581b^2),
  \nonumber
\end{equation}
$b^i = B^i/B^i_{c2}$, and $b^m = B^m/B^m_{c2}$. $B^i_{c2}$ and $D^i$
are the second critical field and the demagnetization factor of the
$i-$th grain, and $D^m$ and $B^m_{c2}$ are the mean values of the
demagnetization factor and the second critical field, respectively.
In the last part of Eq.~(\ref{eq:Bi}), we also replaced $D^i$ and
$b^i$ by their average values $D^m$ and $b^m$ (it is clear, that in
the intermediate range of fields and for temperatures not very close
to $T_c$, $b^i\simeq b^m$).

The advantage to use Gaussian $P^i(B)$ is that the contribution of
the $i-$th line to the total second moment can be obtained
analytically as:\cite{Weber93,Khasanov05-RbOs}
\begin{eqnarray}
\langle \Delta
B^{2}\rangle^{i}&=&(\sigma^i/\gamma_\mu)^2+(B^i-B^m)^2\simeq
\nonumber \\
&&G^2(b^m)[\lambda^i(T)]^{-4}+(B^i-B^m)^2.
 \label{eq:Bi-Gauss}
\end{eqnarray}
By substituting it in Eq.~(\ref{eq:B_tot-full}) one gets:
\begin{equation}
\langle \Delta B^{2}\rangle^{tot} =\frac{\sigma^2}{\gamma_\mu^2}
=\int_T^\infty \omega(t) \langle \Delta B^{2}\rangle dt.
 \label{eq:B_tot-Gauss}
\end{equation}
Later on we are going to use this equation in order to fit the
experimental $\mu$SR data. We want to remind that the total second
moment $\langle \Delta B^{2}\rangle^{tot}$ obtained by means of
Eq.~(\ref{eq:B_tot-Gauss}) is determined by: (i) Function $f(T_c)$
that describes distribution of $\lambda_0^{-2}$ as a function of
$T_c$ [see Eq.~(\ref{eq:BCS-weak-coupled})]. (ii) Function
$\omega(T_c)$ describing distribution of the superconducting
fractions with different $T_c$'s. (iii) Gap to $T_c$ ratio
$R=2\Delta_0/k_BT_c$ [see Eq.~(\ref{eq:BCS-weak-coupled})]. (iv) The
mean values of the demagnetization factor $D^m$ and the reduced
magnetic field $b^m$ [see Eq.~(\ref{eq:Bi}) and
(\ref{eq:Bi-Gauss})].

Note that in real experiments on polycrystalline samples with sharp
transition to the superconducting state, $P(B)$ often obeys Gaussian
distribution (see, {\it e.g.}, Refs.~\onlinecite{Weber93} and
\onlinecite{Pumpin90}). The reason is that the vortex lattice within
the small superconducting grain does not remain regular. It is
distorted due to effects of pinning inside the grain, as well as,
near the edges of the grain due to closeness of vortexes to the
surface. As is shown by Brandt,\cite{Brandt88} even small pinning
leads to substantial smearing of the characteristic features of the
ideal $P(B)$ line and, in a case of intermediate pinning, to the
Gaussian shape of $P(B)$. We should emphasize, however, that even in
a case of distorted vortex lattice, the second moment of $P(B)$ is
still a good measure of $\lambda$.\cite{Brandt88}

\subsection{Magnetic field distribution in a granular
superconductor}\label{subsec:Theoretical background-simulations}

In this section we are going to simulate the internal magnetic field
distribution $P(B)$ within the above described model. As a first
step it is necessary to choose a theoretical model describing the
spatial variation of the local internal magnetic fields in the
vortex lattice [$B({\bf r})$], from which the distribution of the
fields in the $i-$th grain can be obtained as:
\begin{equation}
P^i(B)=\frac{\int\delta(B-B')\ dA(B')}{\int dA(B')}.
\end{equation}
Here, $dA(B')$ is an elemental piece of the vortex lattice unit
cell where the magnetic field is equal to $B'$ and the unit cell
has a total area of $\int dA(B')$. $B({\bf r})$ was calculated by
using an iterative method for solving the Ginzburg- Landau
equations developed by Brandt.\cite{Brandt03} This method allows
to accurately determine $B({\bf r})$ for arbitrary $b$,
$\kappa=\lambda/\xi$ and the vortex lattice symmetry (see
Refs.~\onlinecite{Brandt03}, \onlinecite{Brandt97}, and
\onlinecite{Laulajainen06} for details).
\begin{figure}[htb]
\includegraphics[width=0.65\linewidth]{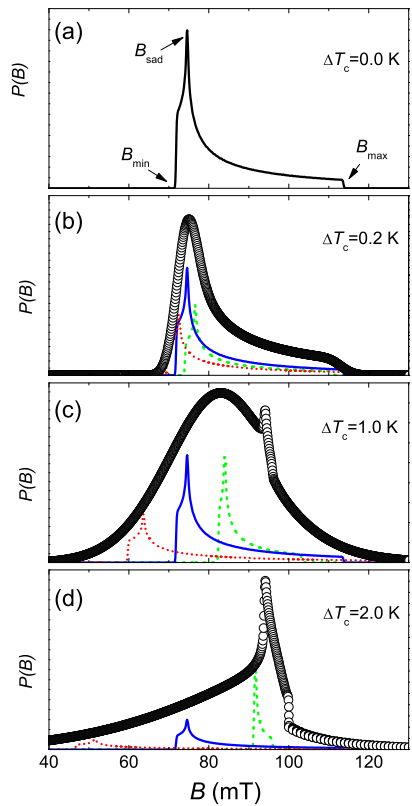}
\caption{(Color online) $P(B)$ distributions calculated by means of
Eq.~(\ref{eq:P(B)-simulations}) at $T_c^m=6$~K,
$\lambda_0(T_c^m)=60$~nm, $\kappa=1.67$, and $B_{ex}=0.1$~T for the
following values of $\Delta T_c$: 0.0~K (a) 0.2~K (b), 1.0~K (c),
and 2~K (d). The dashed green, solid blue, and doted red lines in
(b), (c), and (d) represent the $\omega(T_c^i)P^i(B)$ term for
$T_c^i=T_c^m-\Delta T_c$, $T_c^i=T_c^m$, and $T_c^i=T_c^m+\Delta
T_c$, respectively. See text for details.}
 \label{fig:simulations}
\end{figure}
It was also assumed that the distribution of the superconducting
volume fractions with different $T_c$'s is described by a Gaussian
distribution:
\begin{equation}
\omega(T_c^i)=\frac{1}{\Delta T_c \sqrt{2\pi}}\exp \left(
-\frac{(T_c^i-T_c^m)^2}{2\Delta T_c^2} \right)
 \label{eq:Tc-gauss}
\end{equation}
($T^m_c$ and $\Delta T_c$ are the mean value and the width of the
distribution, respectively), $\lambda_0^{-2}$ follows the power
law:
\begin{equation}
\lambda_0^{-2}(T_c^i) = (K\cdot T_c^i)^n
 \label{eq:power-law}
\end{equation}
with the exponent $n = 2$, and
$\kappa^i=\lambda^i/\xi^i=const$.\cite{comment} As is mentioned
already in the introduction, the power law dependence of
$\lambda_0^{-2}$ on $T_c$ is observed for various
HTS\cite{Uemura89,Uemura91,Zuev05,Liang05,Sonier07} and molecular
superconductors,\cite{Pratt05} as well as obtained theoretically for
materials having 2D or 3D quantum superconductor to insulator
transition.\cite{Kim91,Schneider00,Schneider04,Schneider07}

As initial parameters for calculations we took $T_c^m=6$~K,
$B_{ex}=0.1$~T, $D^m=1/3$, $\lambda_0(T_c^m)=60$~nm and
$\kappa=1.67$. The resulting field distribution $P(B)$ was
obtained as:
\begin{equation}
P(B)=\sum_i^N\omega(T_c^i)P^i(B).
 \label{eq:P(B)-simulations}
\end{equation}
Calculations were done for $N = 60$ in the region $\pm 3\Delta T_c$
around $T_c^m$. Figure~\ref{fig:simulations} shows $P(B)$
distributions for $\Delta T_c = 0.0$~K, 0.2~K, 1.0~K and 2.0~K
obtained by means of Eq.~(\ref{eq:P(B)-simulations}). The lines in
(b), (c), and (d) represent the $\omega(T_c^i)P^i(B)$ term for
$T_c^i=T_c^m-\Delta T_c$ (dashed green line), $T_c^i=T_c^m$ (solid
blue line), and $T_c^i=T_c^m+\Delta T_c$ (dotted red line). It is
seen that the shape of $P(B)$ changes dramatically with increasing
width of the $\omega(T_c)$ distribution. For small $\Delta T_c$
(when transition to the superconducting state is very sharp) $P(B)$
is asymmetric with the highest weight around the point corresponding
to the so called ''saddle point`` field $B_{sad}$ [see
Figs.~\ref{fig:simulations} (a) and (b)]. It is also seen that all
the characteristic features of $P(B)$ at minimum ($B_{min}$),
maximum ($B_{max}$), and ''saddle point`` fields are smeared out
[see Fig.~\ref{fig:simulations}~(b)]. Note that the simulated $P(B)$
presented in Fig.~\ref{fig:simulations}~(b) looks very similar to
what is observed in $\mu$SR experiments on high-quality single
crystals.\cite{Herlach90,Sonier00} With a further increase of
$\Delta T_c$ the $P(B)$ distribution becomes rather symmetric and,
finally, asymmetric again, but now with the maximum weight around
the external field $B_{ex}=0.1$~T and a very long tail at lower
fields. It is interesting to note, that the parts of the sample with
the lowest $T_c$'s are responsible for the peak appearing slightly
below the external field, as is seen in
Fig.~\ref{fig:simulations}~(c) and (d). The reason is the decrease
of $P^i(B)$ width and the shift of $B^i$ towards to $B_{ex}$ with
increasing $\lambda^i$.

To summarize, in Sec.~\ref{sec:theoretical background} we
described the model allowing to obtain the distribution of the
internal magnetic fields in a granular superconducting sample of
moderate quality and calculated the second moment of this
distribution. Within the framework of this model we also simulated
$P(B)$ profiles for materials with different width of the
superconducting transition $\Delta T_c$. It is remarkable, that
already small $\Delta T_c$ leads to smearing of the characteristic
features of $P(B)$ distribution near $B_{min}$, $B_{sad}$, and
$B_{max}$ characteristic fields. Even though this result is quite
predictable, it has an important impact, since previously smearing
of $P(B)$ was ascribed entirely for the effects of pinning (see,
{\it e.g.}, Ref.~\onlinecite{Sonier00}).

\section{Experimental details}\label{sec:experimental}

\begin{figure}[htb]
\includegraphics[width=0.7\linewidth]{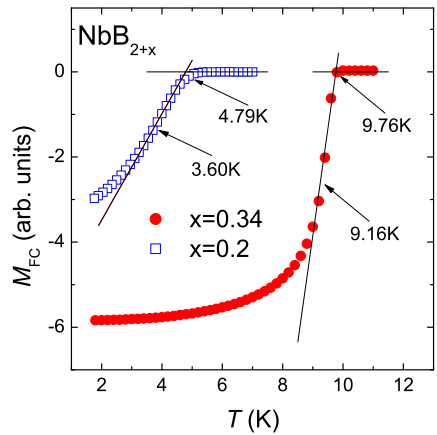}
\caption{(Color online) Temperature dependences of the field-cooled
magnetization ($M_{FC}$) for NbB$_{2.2}$ and NbB$_{2.34}$ samples
measured at $\mu_0H = 0.5$~mT. The solid lines are the best linear
fits to the steepest part of $M_{FC}(T)$ curves and $M = 0$.}
 \label{fig:Magnetization}
\end{figure}

Details of the sample preparation for NbB$_{2+x}$ can be found
elsewhere \cite{Escamilla04}. Both NbB$_{2.2}$ and NbB$_{2.34}$
samples, studied in the present work, were fine powders with the
average grain size of the order of few microns. The field-cooled
0.5~mT magnetization ($M_{FC}$) measurements for NbB$_{2.2}$ and
NbB$_{2.34}$ samples were performed by using a SQUID magnetometer.
The corresponding $M_{FC}(T)$ curves are shown in
Fig.~\ref{fig:Magnetization}. It is seen that the superconducting
transitions are rather broad indicating that both samples are not
particularly uniform. This also implies that the superconducting
critical temperatures may be evaluated only approximately. The
middle-points of transitions correspond to $\simeq9.16$~K and
$\simeq3.60$~K, while linear extrapolations of $M_{FC}(T)$ curves in
the vicinity of $T_c$ to $M = 0$ result in 9.76~K and 4.79~K for
NbB$_{2.2}$ and NbB$_{2.34}$, respectively (see
Fig.~\ref{fig:Magnetization}).

\begin{figure}[htb]
\includegraphics[width=0.8\linewidth]{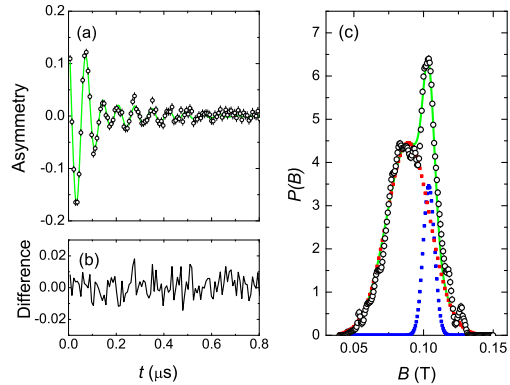}
\caption{(Color online) The muon-time spectra (a), difference
between the two-Gaussian fit and experimental data (b), and internal
field distributions (c) for NbB$_{2.34}$ sample at $T=1.7$~K after
field cooling in magnetic field of 0.1~T. The lines in (a) and (c)
represent the best fit with the Gaussian line-shapes. See text for
details.}
 \label{fig:Fourier}
\end{figure}

The transverse-field $\mu$SR experiments were performed at the
$\pi$M3 beam line at the Paul Scherrer Institute (Villigen,
Switzerland). The powders of NbB$_{2.2}$ and NbB$_{2.34}$ were
cold pressed into pellets (12~mm diameter and 2~mm thick). The
samples were field cooled in $\mu_0H=0.05$~T  (NbB$_{2.2}$) and
0.1~T (NbB$_{2.34}$), applied perpendicular to the flat surface of
the pellet, from above $T_c$ down to 1.7~K. The fields 0.05~T and
0.1~T were chosen in order to perform measurements at the same
reduced magnetic field $B/B_{c2}\simeq0.4$ ($B_{c2}$ is the upper
critical field).\cite{Khasanov06_unp} The $\mu$SR signal was
observed in the usual time-differential way by counting positrons
from decaying muons as a function of time in positron telescopes.
The time dependence of the positron rate is given by the
expression:\cite{msr}
\begin{equation}
 \frac{{\rm d}N(t)}{{\rm d}t} = N_0 {1\over\tau_\mu} e^{-t/\tau_\mu}
  \left[ 1 + A P(t) \right] + bg \; ,
\label{eq:N_t}
\end{equation}
where $N_0$ is the normalization constant, $bg$ denotes the
time-independent background, $\tau_\mu = 2.19703(4) \times
10^{-6}$~s is the muon lifetime, $A$ is the maximum decay asymmetry
for the particular detector telescope ($A\simeq 0.18-0.19$ in our
case), and $P(t)$ is the polarization of the muon ensemble:
\begin{equation}
P(t)=\int P(B)\cos(\gamma_{\mu}Bt+\phi)dB \; .
 \label{eq:P_t}
\end{equation}
Here, $\phi$ is the angle between the initial muon polarization
and the effective symmetry axis of a positron detector. $P(t)$ can
be linked to the internal field distribution $P(B)$ by using the
algorithm of Fourier transform.\cite{msr} The $P(t)$ and $P(B)$
distributions inside the NbB$_{2.34}$ at $T=1.7$~K after
field-cooling in a magnetic field of 0.1~T are shown in
Fig.~\ref{fig:Fourier}~(a) and (c). The $P(B)$ distributions were
obtained from measured $P(t)$ by using the fast Fourier transform
procedure based on the maximum entropy algorithm.\cite{Rainford94}
In order to account for the field distribution seen in
Fig.~\ref{fig:Fourier}~(c), the $\mu$SR time-spectra were fitted
by two Gaussian lines:
\begin{eqnarray}
P(t)=\sum_{i=1}^2A_i \exp(-\sigma_i^2t^2/2) \cos(\gamma_{\mu}B_i
t+\phi) \;.
\label{eq:gauss}
\end{eqnarray}
where $A_i$, $\sigma_i$, and $B_i$ are the asymmetry, the Gaussian
relaxation rate and the first moment of the $i$-th line. At
$T\gtrsim8$~K for NbB$_{2.34}$ and $T\gtrsim4.5$~K for NbB$_{2.2}$
the analysis is simplified to the single line only with
$\sigma_{nm} \simeq 0.3$~$\mu$s$^{-1}$ resulting from the nuclear
moments of the sample. Eq.~(\ref{eq:gauss}) is equivalent to the
field distribution:\cite{Khasanov05-RbOs}
\begin{equation}
P(B)=\gamma_{\mu}\sum_{i=1}^2{A_i \over \sigma_i}
\exp\left(-{\gamma_{\mu}^2(B-B_i)^2 \over 2\sigma_i^2}\right) \; .
\label{eq:P_B}
\end{equation}
The solid line in Fig.~\ref{fig:Fourier}~(a) represent the best fit
with the two-Gaussian lines to the muon-time spectra. The
corresponding $P(B)$ is shown in Fig.~\ref{fig:Fourier}~(c). It
should be mentioned that the two-Gaussian fit can satisfactory
describe the experimental data. For both samples and in the whole
range of temperatures the normalized $\chi^2_{\rm norm}$'s were
found to be close to the unity implying a good quality of fits.

\section{Results and discussions} \label{sec:results_and_discussions}

\begin{figure*}[htb]
\includegraphics[width=0.85\linewidth]{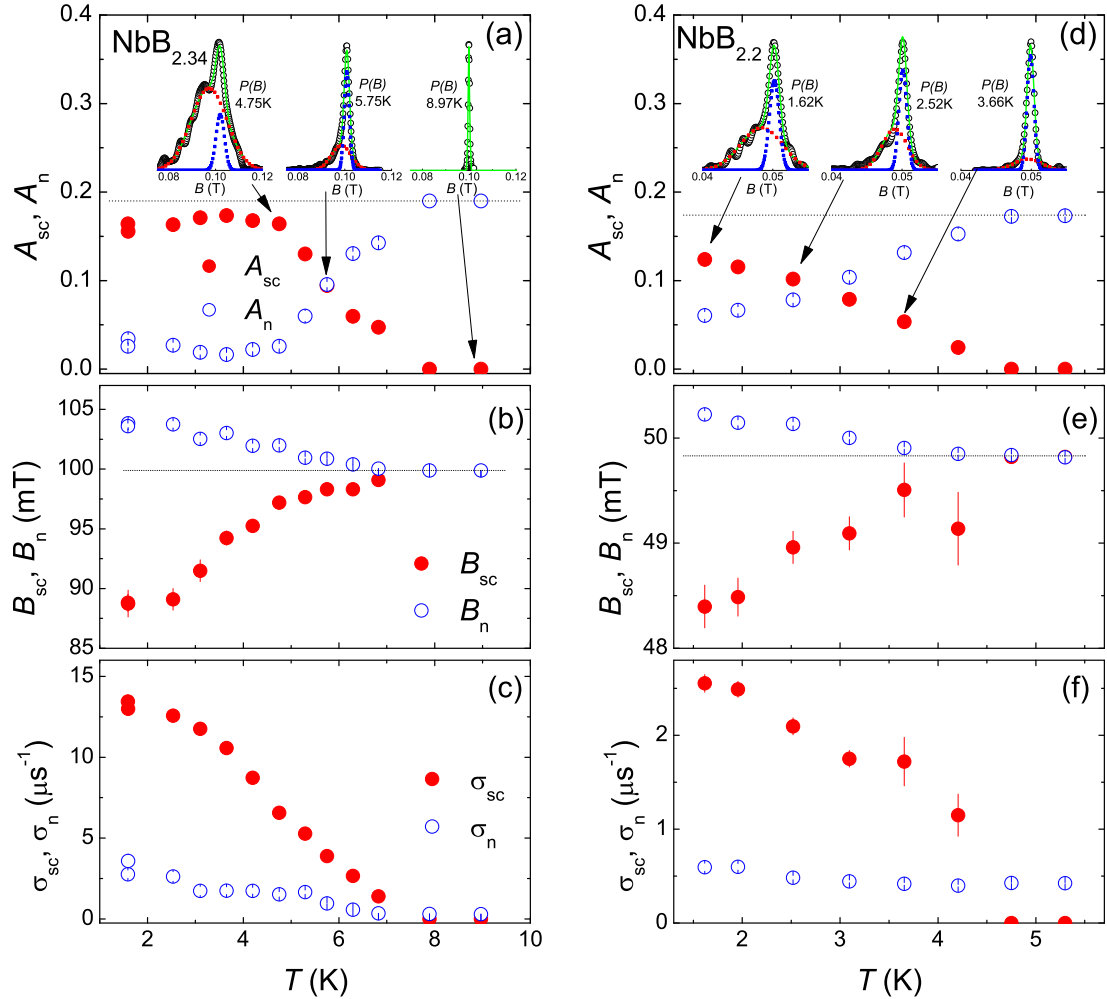}
 \caption{(Color online) The temperature dependences of asymmetries
($A_{sc}$, $A_n$) (a)/(d), internal fields ($B_{sc}$, $B_n$)
(b)/(e), and the muon-spin depolarization rates ($\sigma_{sc}$,
$\sigma_n$) (c)/(f), obtained from the fit of $\mu$SR time spectra
by means of Eq.~(\ref{eq:gauss}) for NbB$_{2.34}$/NbB$_{2.2}$ after
field-cooling in a magnetic field of 0.1~T/0.05~T. The indexes
''sc`` and ''n`` denote the normal and the superconducting state
components. The dotted lines in (a)/(d) and (b)/(e) represent the
total muon asymmetry ($A_{sc}+A_n$) and the external field value
($B_{ex}$), respectively. In the upper panels of (a) and (d) the
$P(B)$ distributions at temperatures marked by the arrows are shown.
The red and the blue dotted lines represent the superconducting
(broad) and the normal state (narrow) components obtained from the
fit by means of Eq.~(\ref{eq:gauss}). The solid green line
corresponds to the sum of these two components.}
 \label{fig:Fourier_Analysis}
\end{figure*}

Results for NbB$_{2.2}$ and NbB$_{2.34}$, obtained from the fit of
experimental data by means of Eq.~(\ref{eq:gauss}), are summarized
in Fig.~\ref{fig:Fourier_Analysis}. Distributions of the local
fields, at temperatures marked by the arrows, are shown in the upper
panels of Figs.~\ref{fig:Fourier_Analysis}~(a) and (d). The broad
lines reflect contributions of the superconducting parts of samples.
Indeed, their asymmetries $A_{sc}$
[Figs.~\ref{fig:Fourier_Analysis}~(a) and (d)] and relaxation rates
$\sigma_{sc}$ [Figs.~\ref{fig:Fourier_Analysis}~(c) and (f)]
increase while first moments $B_{sc}$ [Figs.~(b) and (e)] decrease
with decreasing temperature, as expected for a superconductor in a
mixed state (see, {\it e.g.}, Ref.~\onlinecite{Zimmermann95}). The
narrow lines describe contributions from parts of the samples being
in the normal state. The slight shift of these lines to higher
fields [Figs.~~\ref{fig:Fourier_Analysis}~(b) and (e)] and the small
increase of relaxations [Figs.~~\ref{fig:Fourier_Analysis}~(c) and
(f)] are associated with the diamagnetism of the superconducting
grains leading to increase of the local fields in the
nonsuperconducting parts of the sample. Such behavior is often
observed in samples with less than 100\% superconducting volume
fraction (see, {\it e.g.}, Refs.~\onlinecite{Khasanov05-RbOs} and
\onlinecite{Koda04}).

\begin{figure}[htb]
\includegraphics[width=0.7\linewidth]{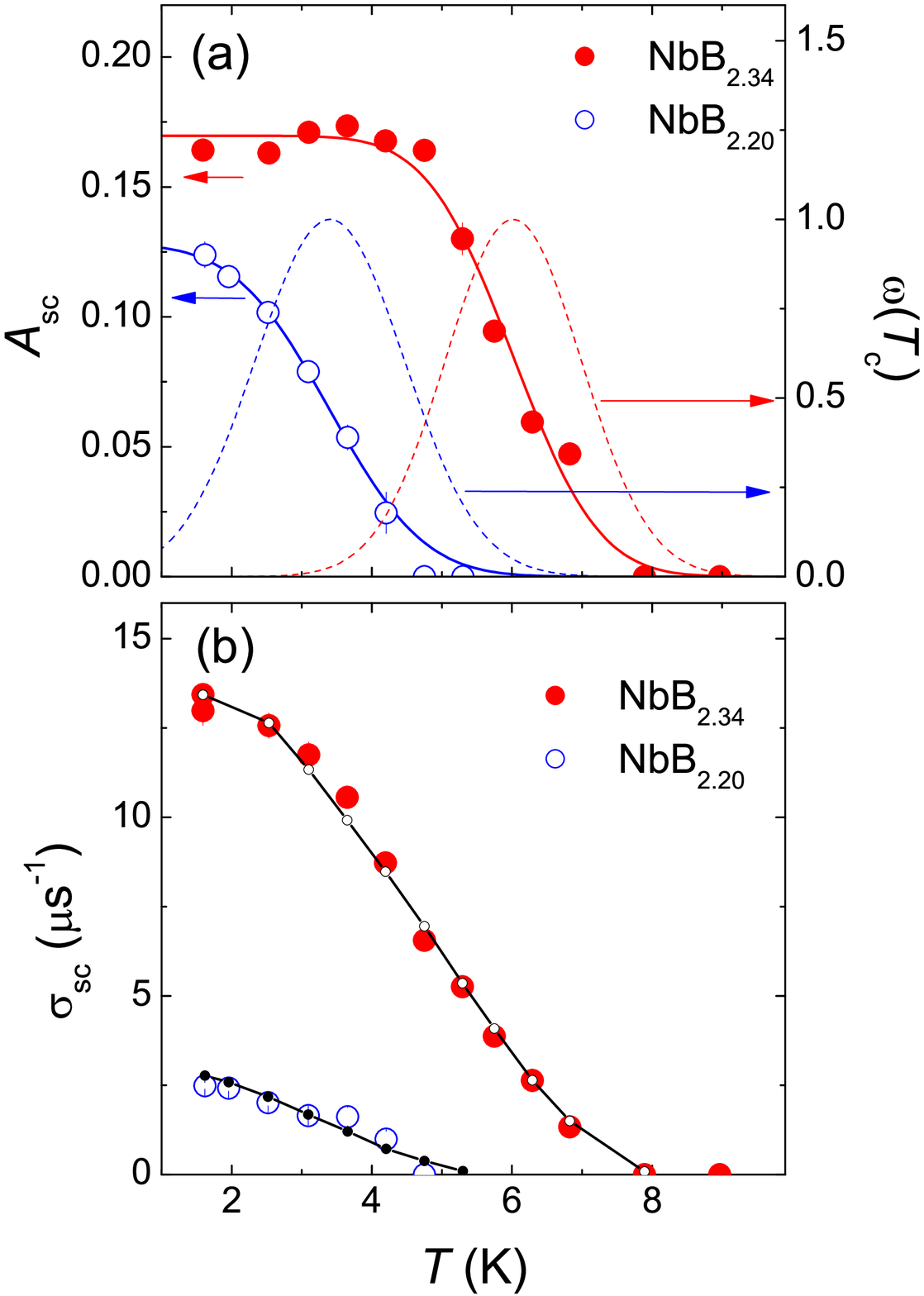}
%
\caption{(Color online) (a) Temperature dependences of the
superconducting asymmetries ($A_{sc}$) for NbB$_{2.34}$ and
NbB$_{2.2}$. The solid lines are the fits of $A_{sc}(T)$ data by
means of Eq.~(\ref{eq:SCfraction-gauss}). The dashed lines
represents the Gaussian distributions of the superconducting volume
fractions with different $T_c$'s [$\omega(T_c)$]. (b) Temperature
dependences of $\sigma_{sc}$ corrected to the nuclear moment
contribution $\sigma_{nm}$ for NbB$_{2.34}$ and NbB$_{2.2}$. The
open and filled small black circles are the fit of $\sigma_{sc}(T)$
by means of Eq.~(\ref{eq:B_tot-Gauss}). See text for details. }
 \label{fig:Asymmetry-Relaxation}
\end{figure}

In Fig.~\ref{fig:Asymmetry-Relaxation} we plot temperature
dependences of the superconducting components $A_{sc}$ and
$\sigma_{sc}$. The $\sigma_{sc}$ values presented in
Fig.~\ref{fig:Asymmetry-Relaxation} are corrected to the nuclear
moment contribution $\sigma_{nm}$ which was subtracted in quadrature
(see {\it e.g.} Ref.~\onlinecite{Khasanov05-RbOs}). It is seen
[Fig.~\ref{fig:Asymmetry-Relaxation}~(a)] that superconductivity
does not disappear abruptly. The superconducting fraction decreases
continuously from their maximum value to zero with temperature
rising from 4~K to 8~K for NbB$_{2.34}$ and from 1.5~K to 5.5~K for
NbB$_{2.2}$. The solid lines in
Fig.~\ref{fig:Asymmetry-Relaxation}~(a) correspond to fits of
$A_{sc}(T)$, assuming that the distribution of the superconducting
volume fractions with different $T_c$'s [$\omega(T_c)$] follows the
Gaussian distribution [see Eq.~(\ref{eq:Tc-gauss})] and,
consequently,
\begin{equation}
A_{sc}(T)=A^m_{sc}\int_T^\infty \omega(t) dt \ .
 \label{eq:SCfraction-gauss}
\end{equation}
Here, $A^m_{sc}$ is the mean value of the superconducting asymmetry
at low temperatures. The fits yield $T^m_{c}$=6.02(3)~K, $\Delta
T_c$=0.96(2)~K, and $A_{sc}^m$=0.170(5) for NbB$_{2.34}$, and
$T^m_{c}$=3.40(4)~K, $\Delta T_c$=1.06(2)~K, and $A^m_{sc}$=0.128(7)
for NbB$_{2.2}$. By taking into account that the total asymmetries
($A_{sc}+A_{n}$) were found to be $\simeq 0.19$ for NbB$_{2.34}$ and
$\simeq0.18$ for NbB$_{2.2}$ [dotted lines in
Figs.~\ref{fig:Fourier_Analysis}(a) and (d)], the superconducting
volume fractions were estimated to be $\simeq$85~\% and
$\simeq$70~\% for NbB$_{2.34}$ and NbB$_{2.2}$, respectively.

The samples used in our $\mu$SR experiments were cold pressed
powders. In this case the individual grains are expected to be only
weakly coupled. This implies that the model, described in
Sec.~\ref{subsec:Theoretical background-calculations}, can be
applied for the particular NbB$_{2.2}$ and NbB$_{2.34}$ samples
studied in the present work. All the variables and functions needed
for such calculations as, {\it e.g.}, the distribution of the
superconducting volume fractions with different $T_c$'s
[$\omega(T_c)$ -- dashed red and blue lines in
Fig.~\ref{fig:Asymmetry-Relaxation}~(a)], the temperature
dependences of the square root of the second moment
[$\sigma_{sc}(T)=\gamma_\mu \sqrt{\langle \Delta B^{2}\rangle(T)}$,
see Figs.~\ref{fig:Fourier_Analysis}~(c) and (f)] and the mean field
[$B_{sc}(T)= B^m(T)$, see Figs.~\ref{fig:Fourier_Analysis}~(b) and
(e)] are directly obtained in the experiments. The only unknown
parameter is the demagnetization factor $D^m$ which enters
Eq.~(\ref{eq:Bi}). We should mention, however, that it is not
correct to assume that $D^m$ is equal to the some average
demagnetization factor for all the grains as, {\it e.g.}, $D^m =
1/3$ for grains of spherical symmetry. The reason for that is the
following. The individual grains have different internal magnetic
fields and, therefore, will show a different diamagnetism. As a
consequence, the internal field in the particular grain is
determined by the diamagnetism of the grain itself, by the
demagnetization field of the whole sample and by the fields from the
other superconducting grains surrounding it. This problem was
already discussed by Weber {\it et al.} in
Ref.~\onlinecite{Weber93}. According to their calculations for the
grains of spherical symmetry the factor $1-D^m$ in Eq.~(\ref{eq:Bi})
should be replaced with [see Eq.~(36) in Ref.~\onlinecite{Weber93}]:
\begin{equation}
1-D^m\simeq2/3-(D^p-1/3)\eta_p/\eta_G.
 \label{eq:1-Dm}
\end{equation}
Here, $D^p$ is the demagnetization factor of the whole sample, and
$\eta_p$ and $\eta_G$ are the effective volume density and the
x-ray density of the sample, respectively. For the experimental
geometry (thin disk in perpendicular magnetic field) $D^p\simeq1$.
Assuming now that the density of the pellet is twice as small as
the x-ray density of the material ($\eta_G/\eta_p= 2$) we get
$1-D^m\simeq0.3$. At this stage we are not going to estimate the
value of $1-D^m$ more precisely. As is shown below, it can be
obtained self consistently from the fit of
Eq.~(\ref{eq:B_tot-Gauss}) to the experimental data.

The demagnetization factor $D^m$ enters Eq.~(\ref{eq:B_tot-Gauss})
via the term $\langle \Delta B^{2}\rangle^i$ that, in its turn, is a
sum of $G^2(b)[\lambda^i(T)]^{-4}$ and $(B^i-B^m)^2$ [see
Eq.~(\ref{eq:Bi-Gauss})]. The term $(B^i-B^m)^2$, which depends on
$D^m$, is responsible for the correction to $\langle \Delta
B^{2}\rangle^i$ appearing due to the shift of the internal field of
the $i-$th grain from the mean internal field $B^m$ [see
Eq.~(\ref{eq:Bi-Gauss})]. It is clear, that during fit of
Eq.~(\ref{eq:B_tot-Gauss}) to the experimental data, contribution of
$(B^i-B^m)^2$ term to $\langle \Delta B^{2}\rangle^{tot}$ needs to
be minimized ($B^m$ should become the first moment of the resulting
$P(B)$ distribution). We used, therefore, an iterative approach.
During the fit, both $\sigma_{sc}(T)$ data sets for NbB$_{2.2}$ and
NbB$_{2.34}$ samples were fitted simultaneously. The dependence of
$\lambda_0$ on the transition temperature was assumed to be
described by the power law $\lambda_0^{-2}=(K\cdot T_c)^n$ [see
Eq.~(\ref{eq:power-law}) and
Refs.~\onlinecite{Zuev05,Liang05,Sonier07,Pratt05,Kim91,
Schneider00,Schneider04,Schneider07}]. In a first step, $1-D^m$ was
assumed to be equal to 0.3 (see above) and
Eq.~(\ref{eq:B_tot-Gauss}) was fitted to the data with the
proportionality factor $K$, the power law exponent $n$, and the gap
to $T_c$ ratio $R=\Delta_0/k_BT_c$ as free parameters. In a second
step, the values of $K$, $n$, and $R$, obtained in the step one,
were substituted back to Eq.~(\ref{eq:B_tot-Gauss}) and the second
term $(B^i-B^m)^2$ entering $\langle \Delta B^{2}\rangle$ in
Eq.~(\ref{eq:B_tot-Gauss}) was minimized with the only free
parameter $D^m$. Then the whole cycle was repeated by using as
initial parameter the newly obtained $D^m$ value. After 3 iterations
the fit already converges. The fit yields $K=
1.24(3)\cdot10^{-2}$~nm$^{-2/n}$K$^{-1}$, $n=3.1(1)$, $R=1.68(3)$
and $1-D^m=0.24(2)$. The open and the filled black circles in
Fig.~\ref{fig:Asymmetry-Relaxation}~(b) represent the result of the
fit of Eq.~\ref{eq:B_tot-Gauss} to the data.

Three important points emerge:\\
(i) The value of $R=\Delta_0/k_BT_c=1.68(3)$ obtained from the fit
is very close to the weak-coupling BCS value 1.76 and
$\Delta_0/k_BT_c=1.55$ obtained by Kotegawa {\it et
al.}\cite{Kotegawa02} in $^{11}$B NMR experiments. It is, however,
much smaller than $\Delta_0/k_BT_c=2.15-2.25$ reported by Ekino
{\it et al.}\cite{Ekino04} in tunneling experiments.\\
(ii) The value of $1-D^m=0.24(2)$ was found to be rather close to
0.3 roughly estimated from Eq.~(\ref{eq:1-Dm}).\\
(iii) For each particular data set the $\lambda_0^{-2}$ {\it vs.}
$T_c$ dependence can be reconstructed {\it only} for $T_c$'s in the
range $T_c^l<T_c<T_c^h$ ($T_c^h$ and $T_c^l$ denote the temperatures
at which the superconducting fraction achieves the maximum value and
vanishes, respectively). Figure~\ref{fig:Asymmetry-Relaxation}~(a)
reveals that the corresponding regions are from $\simeq4$~K to
$\simeq8$~K and from $\simeq1.5$~K to $\simeq5.5$~K for NbB$_{2.34}$
and NbB$_{2.2}$, respectively. Bearing in mind that fit by means of
Eq.~(\ref{eq:B_tot-Gauss}) was performed for both NbB$_{2.34}$ and
NbB$_{2.2}$ data sets simultaneously, we can thus conclude that the
relation $\lambda_0^{-2}\propto T_c^{3.1(1)}$ is valid at least for
temperatures in the region from $\simeq1.5$~K to $\simeq8$~K.

\begin{figure}[htb]
\includegraphics[width=0.7\linewidth]{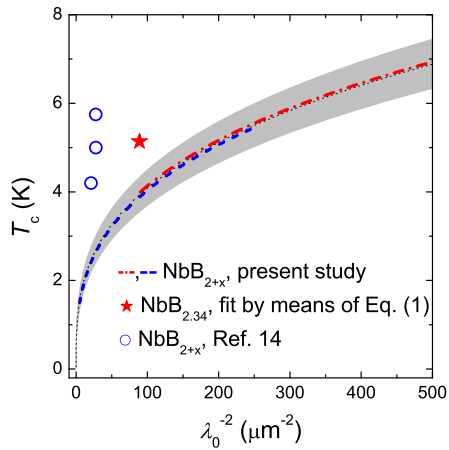}
\caption{ (Color online) Dependence of the transition temperature
($T_c$) on the inverse squared zero-temperature magnetic penetration
depth ($\lambda_0^{-2}$) for NbB$_{2+x}$. The grey area represents
the region where $\lambda_0^{-2}=[(1.24\pm0.03)\cdot10^{-2}\cdot
T_c]^{3.1\pm0.1}$. The dotted line corresponds to
$\lambda_0^{-2}=(1.24\cdot10^{-2}\cdot T_c)^{3.1}$. The dashed blue
and the dash-dotted red lines represent the range of transition
temperatures where $T_c$ {\it vs.} $\lambda^{-2}_0$ dependences were
reconstructed for NbB$_{2.2}$ ($1.5$~K$\lesssim T_c\lesssim5.5$~K)
and NbB$_{2.34}$ ($4$~K$\lesssim T_c\lesssim8$~K). The lines are
shifted by 0.05~K above and below the central
$(1.24\cdot10^{-2}\cdot T_c)^{3.1}$ line for clarity. The star is
the $T_c$ {\it vs.} $\lambda_0^{-2}$ point obtained from the fit of
$\sigma_{sc}(T)$ data for NbB$_{2.2}$ by means of
Eq.~(\ref{eq:BCS-weak-coupled}) when all the measured points are
equally included in the fit. The open circles are the data points
from Ref.~\onlinecite{Takagiwa04}. Values of $\lambda(0)^{-2}$ were
obtained from $\sigma_{sc}(0)$ measured in a field 0.1~T and
$H_{c2}(0)$ (see Ref.~\onlinecite{Takagiwa04}) by using Eq.~(13)
from Ref.~\onlinecite{Brandt03}.}
 \label{fig:Uemura_NbB}
\end{figure}

In Fig.~\ref{fig:Uemura_NbB} we plot $T_c$ {\it vs.}
$\lambda_0^{-2}$ dependence obtained from the fit of  NbB$_{2.34}$
and NbB$_{2.2}$ data. The grey area represents the region where
$\lambda_0^{-2}=[(1.24\pm0.03)\cdot10^{-2}\cdot T_c]^{3.1\pm0.1}$.
The dotted line corresponds to
$\lambda_0^{-2}=[1.24\cdot10^{-2}\cdot T_c]^{3.1}$. The dashed blue
and the dot-dashed red lines represent the range of the transition
temperatures where $T_c$ {\it vs.} $\lambda_0^{-2}$ dependences were
reconstructed for NbB$_{2.2}$ and NbB$_{2.34}$, respectively. For
clarity, these lines are shifted by 0.05~K above and below the
central $(1.24\cdot10^{-2}\cdot T_c)^{3.1}$ line. We also include in
this graph data points for NbB$_{2+x}$ ($x=0.0$, 0.01, and 0.1) from
Ref.~\onlinecite{Takagiwa04}. It is seen that these samples have
approximately twice as high transition temperatures as one would
expect from the obtained $T_c$ {\it vs.} $\lambda_0^{-2}$
dependence. On the base of our results, we can argue that samples
studied in Ref.~\onlinecite{Takagiwa04} also have distributions of
the superconducting volume fractions, similar to what is observed in
the present study. Without taking into account this distributions,
fit of $\sigma_{sc}(T)$ data can lead to a substantial overestimate
of the superconducting transition temperature $T_c$. As an example,
the star in Fig.~\ref{fig:Uemura_NbB} represents result of the fit
of Eq.~(\ref{eq:BCS-weak-coupled}) to $\sigma_{sc}(T)$ NbB$_{2.2}$
data when all measured points are taken equally into account.

Now we are going to comment shortly the observed $T_c$ {\it vs.}
$\lambda_0^{-2}$ dependence. Recently it was shown that the famous
''Uemura`` relation, establishing the linear proportionality between
$T_c$ and $\lambda^{-2}_0$,\cite{Uemura89,Uemura91} does not hold
for highly underdoped HTS's.\cite{Zuev05,Liang05,Sonier07} Indeed,
for YBa$_2$Cu$_3$O$_{7-\delta}$ Zuev {\it et al.} \cite{Zuev05}
observed $\lambda^{-2}_0\propto T_c ^{2.3(4)}$ and found that this
power law is in fairly good agreement with
YBa$_2$Cu$_3$O$_{7-\delta}$ data in the whole doping range. For the
similar compounds Sonier {\it et al.} \cite{Sonier07} obtained
$n=2.6-3.1$. Those values are very close to $3.1(1)$ obtained in the
present study for NbB$_{2+x}$ superconductor. The observation of the
power law type of relation between the transition temperature $T_c$
and the superfluid density in BCS supercondutor NbB$_{2+x}$ and
their good correspondence to what is observed in HTS's points to
close similarity between these materials.
It, probably, comes from the fact that the increase of $T_c$ in
NbB$_{2+x}$, similar to HTS, is determined by increasing the charge
carrier concentration.\cite{Takagiwa04,Escamilla06} Indeed, the
x-ray photoelectron spectroscopy experiments of Escamilla and Huerta
\cite{Escamilla06} show that the increase of boron content leads to
a decrease in the contribution of the Nb $4d$ states and increase in
the contribution of the B $2p_\pi$ states to the density of states
at the Fermi level.

\begin{figure}[htb]
\includegraphics[width=0.7\linewidth]{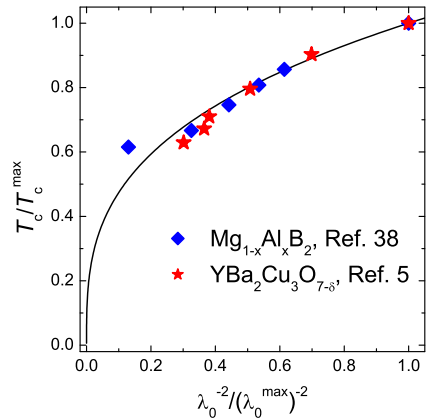}
\caption{ (Color online) Dependence of the transition temperature
normalized to the maximum transition temperature of the
superconducting family ($T_c/T_c^{max}$) on the normalized inverse
squared zero-temperature magnetic penetration depth
[$\lambda_0^{-2}/(\lambda^{max}_0)^{-2}]$. Solid line corresponds to
$\lambda_0^{-2}\propto T_c^{3.1}$ as obtained in the present study.
Blue diamonds are the Mg$_{1-x}$Al$_x$B$_2$ data from
Ref.~\onlinecite{Serventi04}. Red stars are the data points for
YBa$_2$Cu$_3$O$_{7-\delta}$ from Ref.~\onlinecite{Sonier07}. }
 \label{fig:Uemura_NbB_MgB2_Y123}
\end{figure}

It is important to emphasize that observation of correlation between
$T_c$ and $\lambda^{-2}_0$ in BCS superconductors is not restricted
to the particular NbB$_{2+x}$ system studied here. As an example, in
Fig.~\ref{fig:Uemura_NbB_MgB2_Y123} we plot $T_c/T_c^{max}$ as a
function $\lambda_0^{-2}/(\lambda^{max}_0)^{-2}$ for Al doped
MgB$_2$ from Ref.~\onlinecite{Serventi04}. Here $T_c^{max}$ is the
maximum $T_c$ of a certain superconducting family and
$\lambda^{max}_0$ is the corresponding zero temperature penetration
depth. We also include on this graph points for
YBa$_2$Cu$_3$O$_{7-\delta}$ from Ref.~\onlinecite{Sonier07}. It is
seen that all these superconductors represent the very similar
relations.  The scaling relation for the BCS superconductors,
reported here, and their agreement with what was observed in
high-temperature cuprate superconductors
\cite{Zuev05,Liang05,Sonier07} strongly suggests that there are some
features of their electronic properties that are {\it common},
despite these materials have quite different dimensionality, Fermi
surface topology, symmetry of the superconducting order parameter
{\it etc}.

\section{Conclusions} \label{sec:conclusions}

Muon-spin rotation studies were performed on BCS superconductors
NbB$_{2+x}$ ($x=0.2$, 0.34). As a real space microscopic probe,
$\mu$SR allows  to distinguish between the superconducting and
nonsuperconducting parts of the samples and determine the
distributions of the superconducting volume fractions with different
$T_c$'s.  By using the model, developed for a granular
superconductor of moderate quality, the dependence of the
zero-temperature superfluid density $\rho_s\propto \lambda^{-2}_0$
on the transition temperature $T_c$ was reconstructed in a broad
range of temperatures ($1.5$~K$\lesssim T_c\lesssim8.0$~K) revealing
$\rho_s\propto \lambda^{-2}_0\propto T_c^{3.1(1)}$. This dependence
appears to be common at least for some families of BCS
superconductors as, {\it e.g.}, Al doped MgB$_2$ and
high-temperature cuprate superconductors as, {\it e.g.}
YBa$_2$Cu$_3$O$_{7-\delta}$.

\section{Acknowledgments}

This work was partly performed at the Swiss Muon Source (S$\mu$S),
Paul Scherrer Institute (PSI, Switzerland). The authors are
grateful to T.~Schneider for stimulating discussions, and A.~Amato
and D.~Herlach for assistance during the $\mu$SR measurements.
This work was supported by the Swiss National Science Foundation,
by the K.~Alex~M\"uller Foundation, and in part by the SCOPES
grant No. IB7420-110784 and the EU Project CoMePhS.


\begin{thebibliography}{99}
%
\bibitem{Uemura89} Y.J.~Uemura , G.M.~Luke, B.J.~Sternlieb, J.H.~Brewer,
J.F.~Carolan, W.N.~Hardy, R.~Kadono, J.R.~Kempton, R.F.~Kiefl,
S.R.~Kreitzman, P.~Mulhern, T.M.~Riseman, D.Ll.~Williams,
B.X.~Yang, S.~Uchida, H.~Takagi, J.~Gopalakrishnan, A.W.~Sleight,
M.A.~Subramanian, C.L.~Chien, M.Z.~Cieplak, Gang~Xiao, V.Y.~Lee,
B.W.~Statt, C.E.~Stronach, W.J.~Kossler, and X.H.~Yu,
Phys.~Rev.~Lett. {\bf62}, 2317 (1989).
%
\bibitem{Uemura91} Y.J.~Uemura, L.P.~Le, G.M.~Luke, B.J.~Sternlieb,
W.D.~Wu, J.H.~Brewer, T.M.~Riseman, C.L.~Seaman, M.B.~Maple,
M.~Ishikawa, D.G.~Hinks, J.D.~Jorgensen, G.~Saito, and H.~Yamochi,
Phys.~Rev.~Lett. {\bf66}, 2665 (1991).
%
\bibitem{Zuev05} Yu.~Zuev, M.S.~Kim, and T.R.~Lemberger,
Phys.~Rev.~Lett. {\bf 95}, 137002 (2005).
%
\bibitem{Liang05} R.~Liang, D.A.~Bonn, W.N.~Hardy, and D.~Broun,
Phys.~Rev.~Lett. {\bf 94}, 117001 (2005).
%
\bibitem{Sonier07} J.E.~Sonier, S.A.~Sabok-Sayr, F.D.~Callaghan,
C.V.~Kaiser, V.~Pacradouni, J.H.~Brewer, S.L.~Stubbs, W.N.~Hardy,
D.A.~Bonn, R.~Liang, and W.A.~Atkinson, to appear in Phys.~Rev.~B,
arXiv:0706.2882.
%
\bibitem{Pratt05} F.L.~Pratt and S.J.~Blundell, Phys.~Rev.~Lett. {\bf 94}, 097006
(2005).
%
\bibitem{Kim91} K.~Kim and P.B.~Weichman, Phys.~Rev.~B {\bf 43},
13583 (1991).
%
\bibitem{Schneider00} T.~Schneider and J.M.~Singer, {\it Phase
Transition Approach to High Temperature Superconductivity}
(Imperial College Press, London, 2000).
%
\bibitem{Schneider04} T.~Schneider, in {\it The Physics of
Superconductors} (edited by K. H. Bennemann and J. B. Ketterson,
Springer, Berlin, 2004).
%
\bibitem{Schneider07} T.~Schneider, arXiv:cond-mat/0702468.
%
\bibitem{Kotegawa02} H.~Kotegawa, K.~Ishida, Y.~Kitaoka,
T.~Muranaka, N.~Nakagawa, H.~Takagiwa, and J.~Akimitsu, Physica~C
{\bf 378-381}, 25 (2002).
%
\bibitem{Takasaki04} T.~Takasaki, T.~Ekino, H.~Takagiwa, T.~Muranaka,
H.~Fujii, and J.~Akimitsu, Physica~C {\bf 412-414}, 266 (2004).
%
\bibitem{Ekino04} T.~Ekino, T.~Takasaki, H.~Takagiwa, J.~Akimitsu, and H.~Fujii,
Physica C {\bf 408-410}, 828 (2004).
%
\bibitem{Regalado07} E.~Regalado and R.~Escamilla, J.~Phys.:~Condens.~Matter
{\bf 19}, 376209 (2007).
%
\bibitem{Takagiwa04} H.~Takagiwa, S.~Kuroiwa, M.~Yamazawa, J.~Akimitsu, K.~Ohishi,
A.~Koda, W.~Higemoto, and R.~Kadono, J.~Phys.~Soc.~Jpn. {\bf 74},
1386 (2005).
%
\bibitem{Escamilla04} R.~Escamilla, O.~Lovera, T.~Akachi, A.~Duran, R.~Falconi,
F.~Morales, and R.~Escudero, J.~Phys.:~Condens.~Matter {\bf 16},
5979 (2004).
%
\bibitem{Yamamoto01} A.~Yamamoto, C.~Takao, T.~Masui, M.~Izumi, and
S.~Tajima, Physica~C {\bf 383}, 197 (2002).
%
\bibitem{Tinkham75} M.~Tinkham, ''Introduction to
Superconductivity``, {\it Krieger Publishing company, Malabar,
Florida, 1975}.
%
\bibitem{Gonnelli06} R.S.~Gonnelli, D.~Daghero, G.A.~Ummarino,
A.~Calzolari, M.~Tortello, V.A.~Stepanov, N.D.~Zhigadlo,
K.~Rogacki, J.~Karpinski, F.~Bernardini, and S.~Massidda,
Phys.~Rev.~Lett. {\bf 97}, 037001 (2006).
%
\bibitem{Khasanov04} R.~Khasanov, D.G.~Eshchenko, J.~Karpinski,
S.M.~Kazakov, N.D.~Zhigadlo, R.~Br\"utsch, D.~Gavillet,
D.~Di~Castro, A.~Shengelaya, F.~La~Mattina, A.~Maisuradze,
C.~Baines, and H.~Keller, Phys.~Rev.~Lett. {\bf 93}, 157004 (2004).
%
\bibitem{Muhlschlegel59} B.~M\"uhlschlegel, Z.~Phys. {\bf 155},
313 (1959).
%
\bibitem{Brandt03} E.H.~Brandt, Phys.~Rev.~B {\bf 68}, 054506
(2003).
%
\bibitem{Weber93} M.~Weber, A.~Amato, F.N.~Gygax, A.~Schenck,
H.~Maletta, V.N.~Duginov, V.G.~Grebinnik, A.B.~Lazarev,
V.G.~Olshevsky, V.Yu.~Pomjakushin, S.N.~Shilov, V.A.~Zhukov,
B.F.~Kirillov, A.V.~Pirogov, A.N.~Ponomarev, V.G.~Storchak,
S.~Kapusta, and J.~Bock, Phys.~Rev.~B {\bf 48}, 13022 (1993).
%
\bibitem{Khasanov05-RbOs} R.~Khasanov, D.G.~Eshchenko, D.Di~Castro, A.~Shengelaya,
F.La~Mattina, A.~Maisuradze, C.~Baines, H.~Luetkens, J.~Karpinski,
S.M.~Kazakov, and H.~Keller, Phys.~Rev.~B {\bf 72}, 104504 (2005).
%
\bibitem{Pumpin90} B.~P\"umpin, H.~Keller, W.~K\"undig, W.~Odermatt,
I.M.~Savi\'c, J.W.~Schneider, H.~Simmler, P.~Zimmermann,
E.~Kaldis, S.~Rusiecki, Y.~Maeno, and C.~Rossel, Phys.~Rev.~B {\bf
42}, 8019 (1990).
%
\bibitem{Brandt88} E.H.~Brandt, Phys.~Rev.~B {\bf 37}, 2349
(1988).
%
\bibitem{Brandt97} E.H.~Brandt, Phys.~Rev.~Lett. {\bf 78}, 2208 (1997).
%
\bibitem{Laulajainen06} M.~Laulajainen, F.D.~Callaghan,
C.V.~Kaiser, and J.E.~Sonier, Phys.~Rev.~B {\bf 74}, 054511
(2006).
%
\bibitem{comment} For clean superconductor the zero
temperature value of $\xi_0\propto1/\Delta_0\propto1/T_c$. For $n
= 2$ [see Eq. (8)] $\lambda_0^i\propto 1/T_c$ as well leading to
$\kappa^i=\lambda_0^i/\xi_0^i=const$.
%
\bibitem{Herlach90} D.~Herlach, G.~Majer, J.~Major,
J.~Rosenkranz, M.~Schmolz, W.~Schwarz, A.~Seeger, W.~Templ,
E.H.~Brandt, U.~Essmann, K.~Furderer, and M.~Gladisch,
Hyperfine~Interact. {\bf 63}, 41 (1990).
%
\bibitem{Sonier00} J.~Sonier, J.~Brewer, and R. Kiefl, Rew.~Mod.~Phys.
{\bf 72}, 769 (2000).
%
\bibitem{Khasanov06_unp} R.~Khasanov {\it et al.}, unpublished.
%
\bibitem{msr}  A.~Schenck, {\it Muon Spin Rotation: Principles and Applications
in Solid State Physics},  (Adam Hilger, Bristol, 1986);
S.F.J.~Cox, {\sl J.~Phys.} {\bf C20}, 3187 (1987);
 J.H.~Brewer,  ``Muon Spin Rotation/\-Relaxation/\-Resonance''
in {\it Encyclopedia of Applied Physics}  Vol.~11, p.~23 (VCH, New
York, 1995).
%
\bibitem{Rainford94} B.D.~Rainford and G.J.~Daniell, Hyperfine~Interact. {\bf 87},
1129 (1994).
%
\bibitem{Zimmermann95} P.~Zimmermann, H.~Keller, S.~L.~Lee, I.~M.~Savic, M.~Warden,
D.~Zech, R.~Cubitt, E.~M.~Forgan, E.~Kaldis, J.~Karpinski, and
C.~Kr\"uger, Phys.~Rev.~B {\bf 52},  541  (1995).
%
\bibitem{Koda04} A.~Koda, W.~Higemoto, K.~Ohishi, S.R.~Saha, R.~Kadono,
S.~Yonezawa, Y.~Muraoka, and Z.~Hiroi, J.~Phys.~Soc.~Jpn. {\bf
74}, 1678 (2005).
%
%
\bibitem{Escamilla06} R.~Escamilla and L.~Huerta,
Supercond.~Sci.~Technol. {\bf 19}, 623 (2006).
%
\bibitem{Serventi04} S.~Serventi, G.~Allodi, R.De~Renzi, G.~Guidi, L.~Romano, P.~Manfrinetti,
A.~Palenzona, Ch.~Niedermayer, A.~Amato, and Ch.~Baines,
Phys.~Rev.~Lett. {\bf 93}, 217003 (2004).
%
\end{thebibliography}
\end{document}